\documentclass[aip,graphicx]{revtex4-1}

\usepackage{amsmath}
\usepackage{array}
\usepackage{caption}
\usepackage{subcaption}
\usepackage{tikz}
\usepackage{graphicx}
\usepackage{soul}

\begin{document}


\title{A model for the current-voltage characteristic of membrane/electrolyte junctions \footnote{Copyright 2024 L. J. Mele, M. A. Alam and P. Palestri. This article is distributed under a Creative Commons Attribution-NonCommercial-NoDerivs 4.0 International (CC BY-NC-ND) License. The following article has been submitted to Applied Physics Letters journal. After it is published, it will be found at https://pubs.aip.org/aip/apl}} 



\author{L. J. Mele}
\email[]{ljmele@stanford.edu}
\affiliation{Department of Materials Science and Engineering, Stanford University, Stanford, California 94305, USA}

\author{M. A. Alam}
\affiliation{School of Electrical and Computer Engineering, Purdue University, West Lafayette, Indiana 47907, USA}

\author{P. Palestri}
\affiliation{Engineering Department ``Enzo Ferrari", University of Modena and Reggio Emilia, Modena, 41125, Italy}


\date{\today}

\begin{abstract}
A model for the current-voltage characteristic of the junction between an Ion-Sensitive-Membrane and an electrolyte solution is derived and compared with numerical simulations of the Poisson-Nernst-Planck model for ion transport.
The expression resembles that of a semiconductor pn junction with a non-ideality factor of 2. The non-ideality correlated to the voltage drop in the electrolyte induced by the re-arrangement of the counter-ions.
\end{abstract}


\maketitle 

\section{Introduction}

Ion-Sensitive-Membranes (ISM) electrochemical sensors \cite{buhlmann2012} have found broad use in clinical (e.g., blood tests), environmental (e.g., water quality monitoring) and commercial (e.g., food safety) applications \cite{winkler2004, Curulli2021}. ISM sensors operate in open-circuit (zero-current) configuration because the open-circuit voltage, $V_{\rm op}$, is insensitive to parasitic contributions (e.g., series resistance) and it is uniquely related to the analyte concentration by the simple Nernst relationship. In fact, when in contact with an electrolyte (EL), the resulting open-circuit potential is 
\begin{equation*}
    V_{\rm op} = \mathrm{const.} + \frac{k_{\rm B} T}{q}
        \ln \frac{c_{\rm I,B}}{c_{\rm I,ISM}},
\label{eq:Nernst}
\end{equation*}
where $k_{\rm B}$ is the Boltzmann constant, $T$ is the temperature, $q$ is the absolute value of the elementary charge, $c_{\rm I,B}$ is the analyte concentration in the sample electrolyte and $c_{\rm I,ISM}$ is its reference concentration in the ISM. Nernst equation assumes that only specific analyte ions of interest can enter the ISM and compensate the charges associated with ISM doping.

There is, however, an emerging interest in using the steady-state and transient responses of ISM sensors for energy harvesting \cite{Sailapu2021, Ajanta2024, wieczorek2008, slowinska2004, guzinski2015, jin2021, bobacka2001} and characterization of sensor degradation \cite{jin2017}. The results are generally interpreted by phenomenological small circuit equivalent circuits; a physics-based model for steady-state I-V characteristics will create a basis for physical understanding of these small-signal model \cite{sze2006}.

In this letter, we propose a physical model for the current-voltage characteristic of the ISM/EL junction. The structure and the PNP simulations leading to the assumptions use to derive the model and described in Sec.\ref{sec:pnp}.
The derivation of the analytical formula is presented in Sec.\ref{sec:teo}. The formula is then validated by extensive comparison with numerical simulations based on the Poisson-Nernst-Planck model, as shown in Sec.\ref{sec:val}. Our model defines the physical foundation of the phenomenological models used in refs. \cite{Sailapu2021, Ajanta2024}.

\section{PNP-based Numerical modeling of ISM Sensors}
\label{sec:pnp}

\begin{figure}[htbp]
 \centering
 \includegraphics[width=0.6\linewidth]{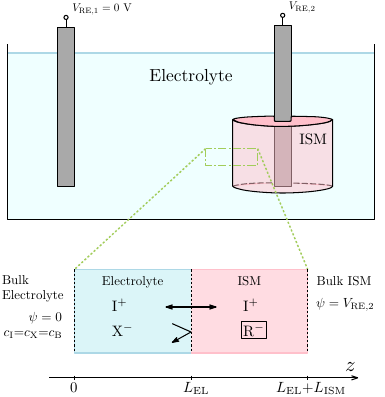}
 \caption{\label{Fig:Sketch}Top plot: sketch of the system under study (not to scale). Bottom panel: 1D domain of the EL/ISM structure considered in this work.}
 \end{figure}

To derive the analytical model, we first define the simulation domain related to problem of interest, see Fig. \ref{Fig:Sketch}.
The simulations assume that a ion $I$ (with valence $z_{\rm I}=+1$ can be exchanged between the membrane (ISM) and the electrolyte (EL), while the counter-ion $X$ (with valence $z_{\rm X}=-1$) is present only in the electrolyte (see Fig. \ref{Fig:Sketch}). This is equivalent to the Donnan exclusion condition \cite{Bergveld1991}.
It is further assumed that the ion $I$ can be exchanged between the ISM and the corresponding contact. In other words, it is assumed that the Redox reactions at the reference electrode in contact with the electrolyte and the ones at the boundary between the membrane and the corresponding electrode are very efficient so that they do not limit the rate of charge transfer that is only limited by phenomena related to the ISM/EL junction. This approximation allows us to derive a model for the ISM/EL junction. This model describes a component of a more complex electrochemical system that may include the finite rate of the Redox reactions mentioned above.
It is also assumed that the ISM is doped with fixed charges with valence $-1$ and concentration $c_{\rm R}$.

Fig. \ref{Fig:Sketch} shows the typical experimental setup of an ISM sensor and one dimensional approximation of the simulation domain. It is implicitly assumed that the curvature of the ISM is small enough to make the radial problem essentially 1D. In addition, the 1D domain places the reference electrode at a distance $L_{\rm EL}$ that is not the distance between the ISM and the actual reference electrode of the 3D domain but, rather, the distance between the ISM and the electrolyte region that is close enough to the reference potential to be considered equipotential.

In the following, we will denote as {\em applied voltage} $V_{\rm App}$ the potential difference applied to the junction in addition to the Nernst potential that builds up due to the difference in ion concentration between ISM and EL.


Since the electrolyte and ISM domains have different chemical properties, the affinity of ionic species can vary greatly. Thus, the Gibbs free energy for ion transfer between the two phases can be more or less favourable. At equilibrium, one can take this into account by using the single-ion distribution coefficient \cite{Amemiya2003} or affinity constant \cite{Mele2022b}, $k_i$, when calculating the Boltzmann ionic distribution. In non-equilibrium conditions, first order forward, $k_i^f$, and backward, $k_i^b$, rates can be defined at the interface \cite{jasielec2012,Mele2022a} such that $k_i=k_i^f/k_i^b$. In this convention, forward (backward) transfer is intended from the  electrolyte (ISM) to the ISM (electrolyte).

\begin{figure}[htbp]
 \centering
 \includegraphics[width=\linewidth]{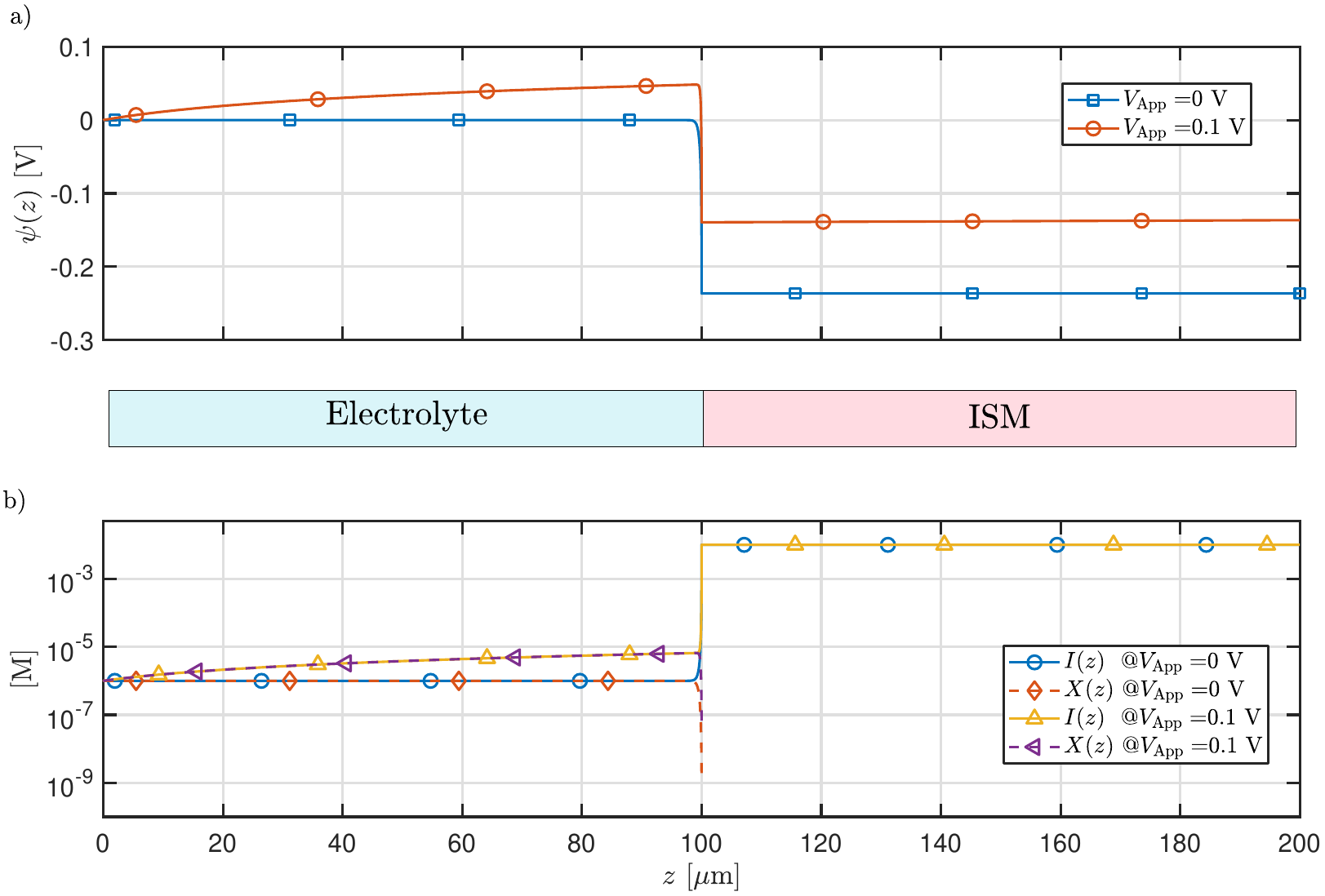}
 \caption{\label{Fig:stacked_profiles}Spatial potential and concentration profiles across the ionic diode structure. (a) Spatial profile of the electrostatic potential at two different applied voltages. $V_{\mathrm{App}}=0$ V also indicates the equilibrium solution or open circuit. (b) Ionic concentration profiles for cations species (solid lines) and anions species (dashed lines) corresponding to the two cases in (a). Parameters values as in Table \ref{tab:Parameters_values_default}.}
 \end{figure}

Sample results of numerical simulations are summarized in Fig. \ref{Fig:stacked_profiles}, and are obtained with the parameters reported in Tab. \ref{tab:Parameters_values_default}. The figure provides several insights:
\begin{itemize}
\item at a short distance from the junction, the concentration of ion $I$ in the ISM (pink region in the bottom panel of Figs. \ref{Fig:Sketch} and \ref{Fig:stacked_profiles}) is equal to the doping $c_{\rm R}$ and the potential is flat (Fig. \ref{Fig:stacked_profiles}(b), ISM domain);
\item the ion $I$ and its counter-ion $X$ have essentially the same concentration profile in the EL (Fig. \ref{Fig:stacked_profiles}(b), electrolyte side), meaning that the electrolyte is essentially neutral;
\item when $V_{\rm App}=0V$, the concentration of $I$ and $X$ in the electrolyte (electrolyte region in Figs. \ref{Fig:Sketch} and \ref{Fig:stacked_profiles}) is flat and equal to the boundary condition set at the reference electrode, denoted as $c_{\rm B}$ (see Fig. \ref{Fig:stacked_profiles}(b)); the potential profile in the electrolyte is flat (see the electrolyte side of Fig. \ref{Fig:stacked_profiles}(a));
\item for $V_{\rm App}\neq 0V$ the concentration profile is essentially linear, ranging from $c_{\rm ITF}$ at the left side of the junction (top panel in Fig. \ref{Fig:stacked_profiles}) to $c_{\rm B}$ at the reference electrode; the potential profile in the electrolyte is not flat (orange curve with circles in Fig. \ref{Fig:stacked_profiles}(a)).
\end{itemize}
The insights from the numerical simulation allows us to derive in the next section an analytical relationship for the I-V characteristics of an ISM sensor.

\begin{table}[htbp]
\centering
\caption{\label{tab:Parameters_values_default}  Simulation parameters for the PNP Model}
\begin{tabular}{|m{8cm}|c|c|}
\hline
\textbf{Parameter} & \textbf{Value} & \textbf{Unit} \\
\hline
\textbf{Diffusion Coefficient (Electrolyte)} & & \\
\hline
- Ion I$^+$, $D_{\mathrm{I}}$ & $1 \times 10^{-9}$ & $\mathrm{m^2/s}$ \\
- Ion X$^-$, $D_{\mathrm{X}}$ & $1 \times 10^{-9}$ & $\mathrm{m^2/s}$ \\
\hline
\textbf{Diffusion Coefficient (ISM)} & & \\
\hline
- Ion I$^+$, $D_{\mathrm{I,ISM}}$ & $1 \times 10^{-11}$ & $\mathrm{m^2/s}$ \\
- Ion X$^-$, $D_{\mathrm{X,ISM}}$ & $1 \times 10^{-11}$ & $\mathrm{m^2/s}$ \\
\hline
\textbf{Interface Rates} & & \\
\hline
- Bulk interfaces, Ion I$^+$, $k_{\mathrm{I}}^{f,b}$ & 100 & $\mathrm{m/s}$ \\
- Bulk interfaces, Ion X$^-$, $k_{\mathrm{X}}^{f,b}$ & 100 & $\mathrm{m/s}$ \\
- Electrolyte/ISM interface, Ion I$^+$, $k_{\mathrm{I}}^{f}$ & 100 & $\mathrm{m/s}$ \\
- Electrolyte/ISM interface, Ion I$^+$, $k_{\mathrm{I}}^{b}$ & 100 & $\mathrm{m/s}$ \\
- Electrolyte/ISM interface, Ion X$^-$, $k_{\mathrm{X}}^{f}$ & 0 & $\mathrm{m/s}$ \\
- Electrolyte/ISM interface, Ion X$^-$, $k_{\mathrm{X}}^{b}$ & 1 & $\mathrm{m/s}$ \\
\hline
\textbf{Simulation Conditions} & & \\
\hline
- Simulation Time & 14400 & s \\
- Electrolyte Layer Thickness, $L_{\rm EL}$ & 100 & $\mu \mathrm{m}$ \\
- ISM Layer Thickness, $L_{\rm ISM}$ & 100 & $\mu \mathrm{m}$ \\
- Electrolyte Bulk Concentration, $c_{\rm B}$ & 1 & $\mu \mathrm{M}$ \\
\hline
\textbf{Fixed Charge (ISM)} & & \\
\hline
- Concentration, $c_{\rm R}$ & 10 & mM \\
- Valence charge, $z_R$ & -1 & \\
\hline
\end{tabular}
\end{table}

\section{Derivation of the Analytical Theory of Current Voltage Characteristics.}
\label{sec:teo}

\subsection{Interface concentration under applied bias}

Since the counter-ion $\rm X$ is at equilibrium (the net flux is zero), the potential drop in the EL must follow Nernst law, so that
\begin{equation}
    \Delta V_{\rm EL}=\frac{k_{\rm B} T}{q}
        \ln \frac{c_{\rm ITF}}{c_{\rm B}}
\label{eq:DV_EL}
\end{equation}
This means that the potential barrier between ISM and EL at the junction is not simply $\Phi_{\rm BI}-V_{\rm App}$ (where $\Phi_{\rm BI}$ is the barrier at $V_{\rm App}=0$ V that builds up due to difference in ion concentration between ISM and EL, i.e., the Donnan potential) but, instead $\Phi_{\rm BI}-V_{\rm App}+\Delta V_{\rm EL}$.
That is the barrier that ions $\rm I^+$ in the ISM should surmount to enter the EL. We can thus write:
\begin{equation}
    c_{\rm ITF}=c_{\rm R} \frac{k_{\rm I}^b}{k_{\rm I}^f}
    \exp \left\{-\frac{k_{\rm B} T}{q} 
    \left( \Phi_{\rm BI}-V_{\rm App}+\Delta V_{\rm EL}
    \right) \right\} 
\label{eq:Citf1}
\end{equation}
Since at $V_{\rm App}=0$ V we have $c_{\rm ITF}=c_{\rm B}$ (that guarantees that the ion fluxes are null), it follows that:
\begin{equation}
    \Phi_{\rm BI}=\frac{k_{\rm B} T}{q}
        \ln \frac{c_{\rm R}k_{\rm I}^f}{c_{\rm B}k_{\rm I}^b}
\label{eq:PhiBI}
\end{equation}
By inserting Eqs. \ref{eq:DV_EL} and \ref{eq:PhiBI} into Eq. \ref{eq:Citf1}, one gets
\begin{equation}
c_{\rm ITF}=c_{\rm B} \exp{  \left( \frac{qV_{\rm App}}{2\,k_{\rm B} T} \right) }.
\label{eq:Citf2} 
\end{equation}
Incidentally, inserting Eq. \ref{eq:Citf2} into Eq. \ref{eq:DV_EL} gives $\Delta V_{\rm EL}=V_{\rm App}/2$, meaning that only half of the applied voltage  contributes in lowering the barrier for ion injection from the ISM into the EL. The other half drops in the electrolyte region.

\subsection{Derivation of the I-V characteristics}

To compute the current, we should consider that only the ion $\rm I^+$ contributes with both drift and diffusion fluxes, while the counter-ion $\rm X^-$ does not carry any current.
We can write the diffusion currents as:
\begin{eqnarray}
J_{\rm I}^{\rm diff} &=& q D_{\rm I} 
    \frac{c_{\rm ITF}-c_{\rm B}}{L_{\rm EL}} \\
J_{\rm X}^{\rm diff} &=& -q D_{\rm X} 
    \frac{c_{\rm ITF}-c_{\rm B}}{L_{\rm EL}}
\end{eqnarray}
where we have exploited the fact that the concentration profiles are essentially linear and equal for both $\rm I^+$ and $\rm X^-$ species. $L_{\rm EL}$ is the length of the EL region, while $D_{\rm X,I}$ are the diffusion coefficients of the ions in the electrolyte.
As for the drift contributions:
\begin{eqnarray}
J_{\rm I}^{\rm drift} &=& q \mu_{\rm I} 
    c (z) F(z) \\
J_{\rm X}^{\rm drift} &=& q \mu_{\rm X} 
    c (z) F(z)
\end{eqnarray}
where the current can be evaluated at each position $z$ and $F(z)$ is the electric field.
Since the current of the $\rm X^-$ ions is null, we have 
\begin{equation}
J_{\rm X}^{\rm diff}=-J_{\rm X}^{\rm drif}
\end{equation}
But, since:
\begin{eqnarray}
J_{\rm I}^{\rm drift} &=& \frac{D_{\rm I}}{D_{\rm X}} J_{\rm X}^{\rm drift} \\
J_{\rm I}^{\rm diff} &=& -\frac{D_{\rm I}}{D_{\rm X}} J_{\rm X}^{\rm diff} 
\end{eqnarray}
it follows that $J_{\rm I}^{\rm diff}=J_{\rm I}^{\rm drift}$
\begin{equation}
J=J_{\rm I}^{\rm drift}+J_{\rm I}^{\rm diff}=2 J_{\rm I}^{\rm diff}=2 q D_{\rm I} 
    \frac{c_{\rm ITF}-c_{\rm B}}{L_{\rm EL}}
\label{eq:J}
\end{equation}
That is, the current is twice the diffusion contribution, similarly to what has been found for quantum-well lasers in \cite{Alam1994}.

Inserting Eq. \ref{eq:Citf2} into Eq. \ref{eq:J} one finally gets:
\begin{equation}
J=\frac{2 q D_{\rm I} c_{\rm B}}{L_{\rm EL}}
 \left[ \exp{ \left( \frac{qV_{\rm App}}{2k_{\rm B} T}\right) } -1 \right]
\label{eq:J_Vapp}
\end{equation}
This is similar to the well-known expression for a semiconductor junction except for the $2$ at the prefactor and the $2k_{\rm B} T$ replacing $k_{\rm B} T$.

The model for the ISM/EL junctions should be complemented by a series resistance representing the voltage drop on the ISM due to the ions $\rm I^+$ carrying the current:
\begin{equation}
    R_{\rm S}= \frac{L_{\rm ISM}}{q \mu_{\rm I,ISM} c_{\rm R} A}
\label{eq:Rseries}
\end{equation}
where $A$ is the area of the junction, $L_{\rm ISM}$ the thickness of the ISM and $\mu_{\rm I,ISM}$ the mobility of the ions in the ISM. 

The final model equation (including the series resistance) can obtained by re-writing Eq. \ref{eq:J_Vapp} as:
\begin{equation}
J=\frac{2 q D_{\rm I} c_{\rm B}}{L_{\rm EL}}
\left[ \exp{ \left( q\frac{V_{\rm App}-J \cdot A \cdot R_{\rm S}}{2k_{\rm B} T} \right)} -1 
\right]
\label{eq:J_Vapp_RS}
\end{equation}
which is an implicit relation between $J$ and $V_{\rm App}$.

\subsection{Small signal equivalent circuit}

Linearization of Eq. \ref{eq:J_Vapp} around the bias $V_{\rm App}$ gives the equivalent resistance
\begin{equation}
    R_{\rm jun}=\frac{1}{A \partial J / \partial V_{\rm App}}=
    \frac{k_{\rm B} T}{q^2 D_{\rm I} c_{\rm B}} \frac{L_{\rm EL}}{A}
    \exp{ \left( -\frac{qV_{\rm App}}{2k_{\rm B} T} \right) }
        = \frac{L_{\rm EL}}{A q \mu_{\rm I} c_{\rm B}}
        \exp{ \left( -\frac{qV_{\rm App}}{2k_{\rm B} T} \right) }.
        \label{eq:Rjun}
\end{equation}
For the bias $V_{\rm App}=0$ V it reduces to the equivalent resistance
 of the electrolyte region. 
 In general, Eq. \ref{eq:Rjun} suggests that the resistance is voltage-dependent, as expected.

 It is important to note that the resistance from Eq. \ref{eq:Rjun} results in series with the resistance $R_{\rm S}$ given by Eq. \ref{eq:Rseries} and with other resistive terms not included in this analysis, such as the contact resistance and the resistance of the portion of electrolyte not included in the 1D domain. With the calculated resistance, one can further estimate the capacitance of the ionic junction, using electrical impedance spectroscopy or charge control experiments \cite{bobacka2001, slowinska2004,Ajanta2024}.

\section{Validation of the analytical model}
\label{sec:val}

The simple expression given by Eq. \ref{eq:J_Vapp} (corrected by the series resistance as in Eq. \ref{eq:J_Vapp_RS} has been extensively compared against numerical simulations.

\begin{figure}[htbp]
\centering
 \includegraphics[width=\linewidth]{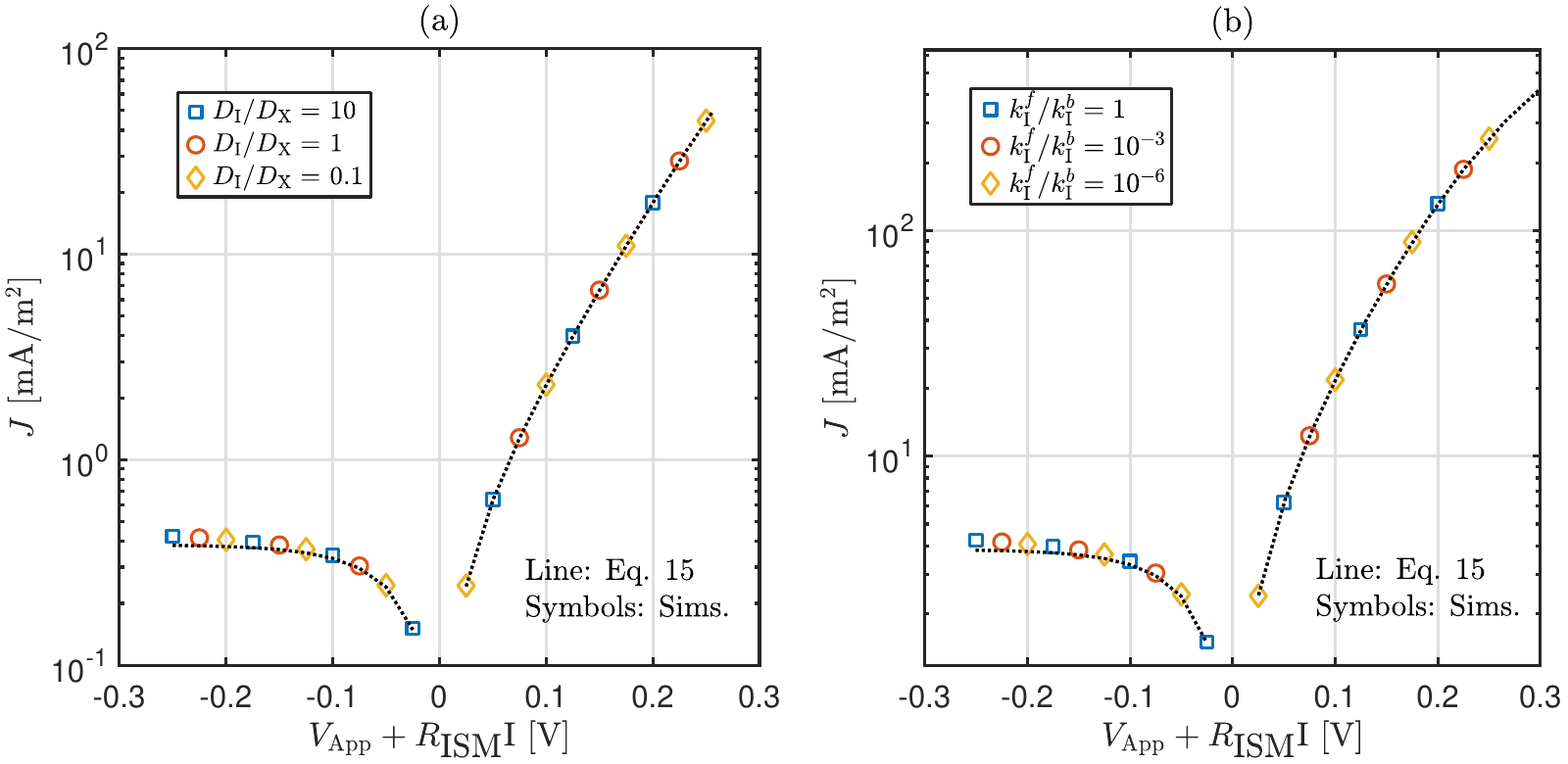}
\caption{I-V characteristic curve for different ratios of diffusion coefficients, (a), and for different ratios of the affinity constant of ion I, (b). Reference parameters values are in Table \ref{tab:Parameters_values_default}.}
\label{fig:DI_and_kI_dependence}
\end{figure}

Fig. \ref{fig:DI_and_kI_dependence} shows that the simulations depends neither on the diffusivity of the $\rm X^-$ ion (Fig.\ref{fig:DI_and_kI_dependence}(a)) nor in the ratio $k_{\rm I}^f/k_{\rm I}^b$ (Fig. 
\ref{fig:DI_and_kI_dependence}(b)), consistently with Eq. \ref{eq:J_Vapp}. The simulations and the equation are in excellent mutual agreement.

\begin{figure}[htbp]
\centering
\includegraphics[width=\linewidth]{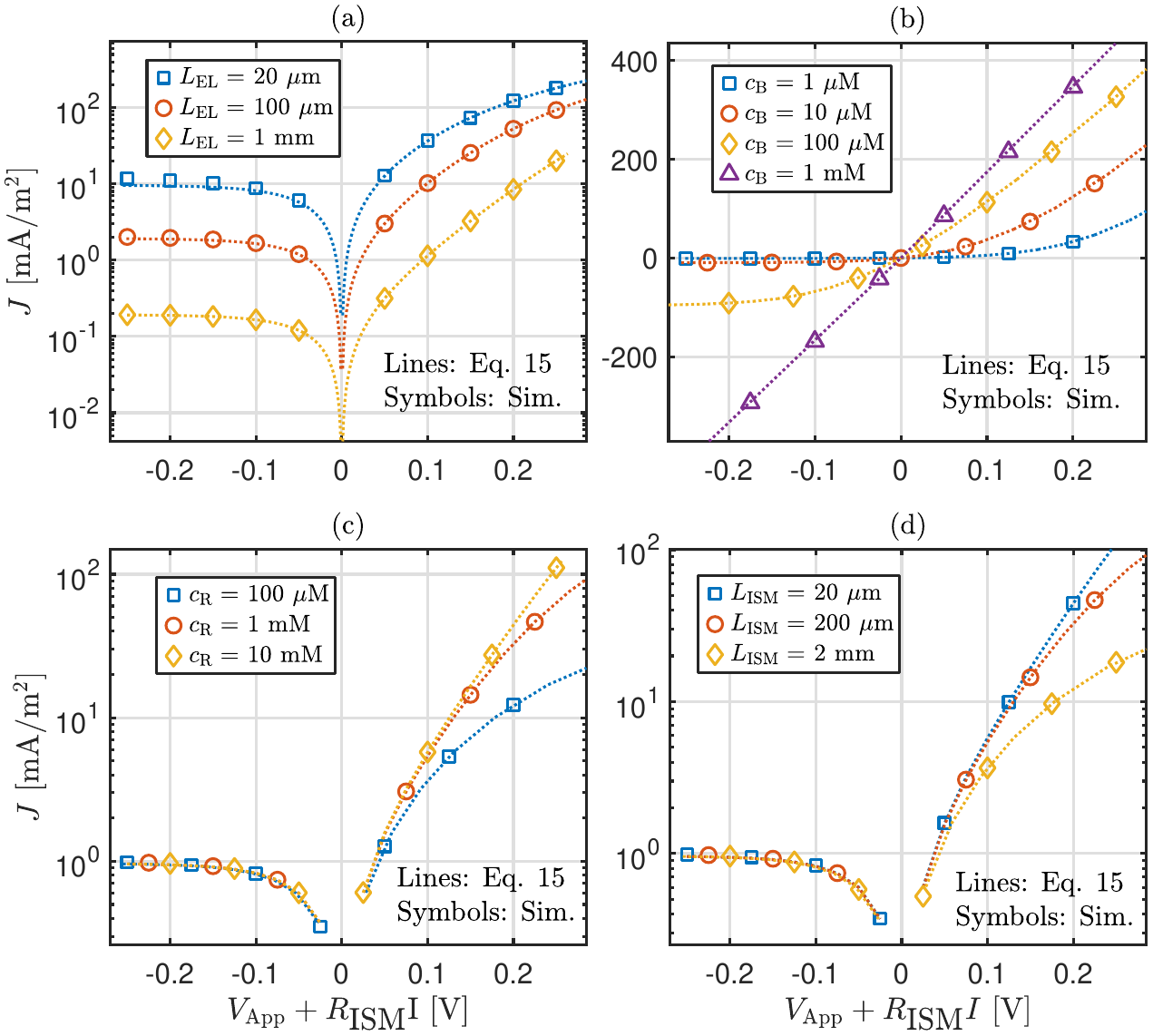}\vspace{-0.25cm}
\caption{I-V characteristic curve for different values of main parameter values: electrolyte thickness, (a), ISM thickness, (b), ionic sites concentration, (c), and bulk electrolyte concentration, (d). PNP simulations (symbols) are compared with Eq. \ref{eq:J_Vapp_RS} (dotted lines). Reference parameters values are in Table \ref{tab:Parameters_values_default}.}
\label{fig:del_dm_cR_cIB_dependence}
\end{figure}

Furthermore, Figs. \ref{fig:del_dm_cR_cIB_dependence}(a,b) show that Eq. \ref{eq:J_Vapp} correctly reproduces the impact of $L_{\rm EL}$ and $c_{\rm B}$, respectively. We also observe that the effect of $c_{\rm R}$ and $L_{\rm ISM}$ (not present in Eq. \ref{eq:J_Vapp}) is to modulate the series resistance (given by Eq. \ref{eq:Rseries} and included in Eq. \ref{eq:J_Vapp_RS}) as shown clearly by Figs. \ref{fig:del_dm_cR_cIB_dependence}(c,d), respectively.
On the other hand, $c_{\rm B}$ has a significant impact in the current: the I-V curves increase linearly with the $c_{\rm B}$ value until the current becomes high enough to make the series resistance dominate (triangles in Fig. \ref{fig:del_dm_cR_cIB_dependence}(b)).

\section{Discussion}

In conclusion, we have derived and validated a compact expression for the static current-voltage characteristic of the ISM/electrolyte junction. This can be seen as an equivalent circuit consisting of a diode with non-ideality factor equal to 2 in series with a resistance due to ion transport in the bulk of the ISM and an ideal voltage generator with value $\Phi_{\rm BI}$. This equivalent circuit can be plug into the equivalent circuit of realistic electrochemical systems. For example if the ISM lays on a blocking interface (e.g., the gate oxide of a MOSFET as in \cite{Mele2022a}) the blocking interface will results in a capacitor in series with the equivalent circuit discussed above representing the ISM/EL junction.
If the redox reactions at the reference electrode or at the interface between ISM and electrode are slow, their I-V model \cite{Jin2020} should be added in series to the model for the EL/ISM junction proposed here.
The impact of slower redox on the overall current-voltage curve should be taken into account when comparing the model proposed here with experimental data. For example, studies on energy harvesters based on ISM and EIS techniques \cite{Sailapu2021, bobacka2001, Ajanta2024} could be used to measure some of the dependencies highlighted in Fig. \ref{fig:del_dm_cR_cIB_dependence}, providing experimental validation of our predicted I-V curve.

The modeling framework proposed here, being based on diffusion of ions injected from an ISM into an electrolyte, can also be used to model and understand the electrodiffusion processes exploited in other studies, such us for measuring the double layer capacitance through applied one-dimensional ionic fluxes \cite{slowinska2004} as well as to derive ionic partitioning coefficients from transient ISM-conditioning measurements \cite{wieczorek2008}.

There are limitations in this derivation that considers only a 1:1 electrolyte without interfering ions. Extensions to multi-ions electrolyte with arbitrary valence is beyond the aim of the present work and will be considered in future publications. Another limitation is that the carrier velocity at the junction is assumed to be limited only by the diffusion velocity ($D_{\rm I}/L_{\rm EL}$ and this may be not accurate when the transfer rates $k_{\rm I}^f$ and $k_{\rm I}^b$ are sufficiently small. However, we verified by simulations based on the PNP model that this happens only when such numbers are unrealistically small.

One should note that the 1D analysis proposed here is not per se a significant limitation. In fact, in realistic devices such as the one sketched in Fig. \ref{Fig:Sketch}, the thickness of the ISM is much smaller than the support electrode size leading to small curvatures where electrodiffusion processes can be effectively assumed one-dimensional. However, when using our model to compare with experimental data, the determination of the parameter $L_{\rm EL}$ should match with the condition of equilibrium dictated case-by-case by the specific geometry. In other words, $L_{\rm EL}$ should be used as fitting parameter in Eq. \ref{eq:J_Vapp_RS}. Finally, the framework employed here can also be used to derive a dynamic model including the junction capacitance and the capacitive contributions related to the finite ion diffusion time in the electrolyte.


\bibliography{bibliography.bib}

\begin{thebibliography}{19}%
\makeatletter
\providecommand \@ifxundefined [1]{%
 \@ifx{#1\undefined}
}%
\providecommand \@ifnum [1]{%
 \ifnum #1\expandafter \@firstoftwo
 \else \expandafter \@secondoftwo
 \fi
}%
\providecommand \@ifx [1]{%
 \ifx #1\expandafter \@firstoftwo
 \else \expandafter \@secondoftwo
 \fi
}%
\providecommand \natexlab [1]{#1}%
\providecommand \enquote  [1]{``#1''}%
\providecommand \bibnamefont  [1]{#1}%
\providecommand \bibfnamefont [1]{#1}%
\providecommand \citenamefont [1]{#1}%
\providecommand \href@noop [0]{\@secondoftwo}%
\providecommand \href [0]{\begingroup \@sanitize@url \@href}%
\providecommand \@href[1]{\@@startlink{#1}\@@href}%
\providecommand \@@href[1]{\endgroup#1\@@endlink}%
\providecommand \@sanitize@url [0]{\catcode `\\12\catcode `\$12\catcode `\&12\catcode `\#12\catcode `\^12\catcode `\_12\catcode `\%12\relax}%
\providecommand \@@startlink[1]{}%
\providecommand \@@endlink[0]{}%
\providecommand \url  [0]{\begingroup\@sanitize@url \@url }%
\providecommand \@url [1]{\endgroup\@href {#1}{\urlprefix }}%
\providecommand \urlprefix  [0]{URL }%
\providecommand \Eprint [0]{\href }%
\providecommand \doibase [0]{http://dx.doi.org/}%
\providecommand \selectlanguage [0]{\@gobble}%
\providecommand \bibinfo  [0]{\@secondoftwo}%
\providecommand \bibfield  [0]{\@secondoftwo}%
\providecommand \translation [1]{[#1]}%
\providecommand \BibitemOpen [0]{}%
\providecommand \bibitemStop [0]{}%
\providecommand \bibitemNoStop [0]{.\EOS\space}%
\providecommand \EOS [0]{\spacefactor3000\relax}%
\providecommand \BibitemShut  [1]{\csname bibitem#1\endcsname}%
\let\auto@bib@innerbib\@empty
\bibitem [{\citenamefont {B{\"u}hlmann}\ and\ \citenamefont {Chen}(2012)}]{buhlmann2012}%
  \BibitemOpen
  \bibfield  {author} {\bibinfo {author} {\bibfnamefont {P.}~\bibnamefont {B{\"u}hlmann}}\ and\ \bibinfo {author} {\bibfnamefont {L.~D.}\ \bibnamefont {Chen}},\ }\bibfield  {title} {\enquote {\bibinfo {title} {Ion-selective electrodes with ionophore-doped sensing membranes},}\ }\href {\doibase 10.1002/9780470661345.smc097} {\bibfield  {journal} {\bibinfo  {journal} {Supramolecular Chemistry: From Molecules to Nanomaterials}\ } (\bibinfo {year} {2012}),\ 10.1002/9780470661345.smc097},\ \bibinfo {note} {doi: \href{https://doi.org/10.1002/9780470661345.smc097}{10.1002/9780470661345.smc097}}\BibitemShut {NoStop}%
\bibitem [{\citenamefont {Winkler}\ \emph {et~al.}(2004)\citenamefont {Winkler}, \citenamefont {Rieger}, \citenamefont {Saracevic}, \citenamefont {Pressl},\ and\ \citenamefont {Gruber}}]{winkler2004}%
  \BibitemOpen
  \bibfield  {author} {\bibinfo {author} {\bibfnamefont {S.}~\bibnamefont {Winkler}}, \bibinfo {author} {\bibfnamefont {L.}~\bibnamefont {Rieger}}, \bibinfo {author} {\bibfnamefont {E.}~\bibnamefont {Saracevic}}, \bibinfo {author} {\bibfnamefont {A.}~\bibnamefont {Pressl}}, \ and\ \bibinfo {author} {\bibfnamefont {G.}~\bibnamefont {Gruber}},\ }\bibfield  {title} {\enquote {\bibinfo {title} {Application of ion-sensitive sensors in water quality monitoring},}\ }\href {\doibase 10.2166/wst.2004.0678} {\bibfield  {journal} {\bibinfo  {journal} {Water Science and Technology}\ }\textbf {\bibinfo {volume} {50}},\ \bibinfo {pages} {105--114} (\bibinfo {year} {2004})}\BibitemShut {NoStop}%
\bibitem [{\citenamefont {Curulli}(2021)}]{Curulli2021}%
  \BibitemOpen
  \bibfield  {author} {\bibinfo {author} {\bibfnamefont {A.}~\bibnamefont {Curulli}},\ }\bibfield  {title} {\enquote {\bibinfo {title} {Electrochemical biosensors in food safety: Challenges and perspectives},}\ }\href {\doibase 10.3390/molecules26102940} {\bibfield  {journal} {\bibinfo  {journal} {Molecules}\ }\textbf {\bibinfo {volume} {26}} (\bibinfo {year} {2021}),\ 10.3390/molecules26102940}\BibitemShut {NoStop}%
\bibitem [{\citenamefont {Sailapu}, \citenamefont {Sabaté},\ and\ \citenamefont {Bakker}(2021)}]{Sailapu2021}%
  \BibitemOpen
  \bibfield  {author} {\bibinfo {author} {\bibfnamefont {S.~K.}\ \bibnamefont {Sailapu}}, \bibinfo {author} {\bibfnamefont {N.}~\bibnamefont {Sabaté}}, \ and\ \bibinfo {author} {\bibfnamefont {E.}~\bibnamefont {Bakker}},\ }\bibfield  {title} {\enquote {\bibinfo {title} {Self-powered potentiometric sensors with memory},}\ }\href {\doibase 10.1021/acssensors.1c01273} {\bibfield  {journal} {\bibinfo  {journal} {ACS Sensors}\ }\textbf {\bibinfo {volume} {6}},\ \bibinfo {pages} {3650--3656} (\bibinfo {year} {2021})}\BibitemShut {NoStop}%
\bibitem [{\citenamefont {Saha}\ and\ \citenamefont {Alam}(2024)}]{Ajanta2024}%
  \BibitemOpen
  \bibfield  {author} {\bibinfo {author} {\bibfnamefont {A.}~\bibnamefont {Saha}}\ and\ \bibinfo {author} {\bibfnamefont {M.~A.}\ \bibnamefont {Alam}},\ }\bibfield  {title} {\enquote {\bibinfo {title} {Signal as an energy source: Dual-functional potentiometric sensor serves as an energy harvester},}\ }\href {\doibase 10.1021/acsaelm.4c00064} {\bibfield  {journal} {\bibinfo  {journal} {ACS Applied Electronic Materials}\ }\textbf {\bibinfo {volume} {6}},\ \bibinfo {pages} {4855--4863} (\bibinfo {year} {2024})}\BibitemShut {NoStop}%
\bibitem [{\citenamefont {Wieczorek}\ and\ \citenamefont {Opalski}(2008)}]{wieczorek2008}%
  \BibitemOpen
  \bibfield  {author} {\bibinfo {author} {\bibfnamefont {P.~Z.}\ \bibnamefont {Wieczorek}}\ and\ \bibinfo {author} {\bibfnamefont {L.~J.}\ \bibnamefont {Opalski}},\ }\bibfield  {title} {\enquote {\bibinfo {title} {{An empirical study of transient responses of potentiometric ion sensors}},}\ }in\ \href {\doibase 10.1117/12.817960} {\emph {\bibinfo {booktitle} {Photonics Applications in Astronomy, Communications, Industry, and High-Energy Physics Experiments 2008}}},\ Vol.\ \bibinfo {volume} {7124},\ \bibinfo {editor} {edited by\ \bibinfo {editor} {\bibfnamefont {R.~S.}\ \bibnamefont {Romaniuk}}\ and\ \bibinfo {editor} {\bibfnamefont {T.~R.}\ \bibnamefont {Wolinski}}},\ \bibinfo {organization} {International Society for Optics and Photonics}\ (\bibinfo  {publisher} {SPIE},\ \bibinfo {year} {2008})\ p.\ \bibinfo {pages} {71240V}\BibitemShut {NoStop}%
\bibitem [{\citenamefont {Slowinska}\ and\ \citenamefont {Majda}(2004)}]{slowinska2004}%
  \BibitemOpen
  \bibfield  {author} {\bibinfo {author} {\bibfnamefont {K.}~\bibnamefont {Slowinska}}\ and\ \bibinfo {author} {\bibfnamefont {M.}~\bibnamefont {Majda}},\ }\bibfield  {title} {\enquote {\bibinfo {title} {{Measurements of the capacitance and the response time of solid-state potentiometric sensors by an electrochemical time-of-flight method}},}\ }\href {\doibase 10.1007/s10008-004-0543-8} {\bibfield  {journal} {\bibinfo  {journal} {Journal of Solid State Electrochemistry}\ }\textbf {\bibinfo {volume} {8}},\ \bibinfo {pages} {763--771} (\bibinfo {year} {2004})}\BibitemShut {NoStop}%
\bibitem [{\citenamefont {Guzinski}\ \emph {et~al.}(2015)\citenamefont {Guzinski}, \citenamefont {Jarvis}, \citenamefont {Pendley},\ and\ \citenamefont {Lindner}}]{guzinski2015}%
  \BibitemOpen
  \bibfield  {author} {\bibinfo {author} {\bibfnamefont {M.}~\bibnamefont {Guzinski}}, \bibinfo {author} {\bibfnamefont {J.~M.}\ \bibnamefont {Jarvis}}, \bibinfo {author} {\bibfnamefont {B.~D.}\ \bibnamefont {Pendley}}, \ and\ \bibinfo {author} {\bibfnamefont {E.}~\bibnamefont {Lindner}},\ }\bibfield  {title} {\enquote {\bibinfo {title} {Equilibration time of solid contact ion-selective electrodes},}\ }\href {\doibase 10.1021/acs.analchem.5b00775} {\bibfield  {journal} {\bibinfo  {journal} {Analytical Chemistry}\ }\textbf {\bibinfo {volume} {87}},\ \bibinfo {pages} {6654--6659} (\bibinfo {year} {2015})}\BibitemShut {NoStop}%
\bibitem [{\citenamefont {Jin}\ \emph {et~al.}(2021)\citenamefont {Jin}, \citenamefont {Saha}, \citenamefont {Jiang}, \citenamefont {Oduncu}, \citenamefont {Yang}, \citenamefont {Sedaghat}, \citenamefont {Maize}, \citenamefont {Allebach}, \citenamefont {Shakouri}, \citenamefont {Glassmaker} \emph {et~al.}}]{jin2021}%
  \BibitemOpen
  \bibfield  {author} {\bibinfo {author} {\bibfnamefont {X.}~\bibnamefont {Jin}}, \bibinfo {author} {\bibfnamefont {A.}~\bibnamefont {Saha}}, \bibinfo {author} {\bibfnamefont {H.}~\bibnamefont {Jiang}}, \bibinfo {author} {\bibfnamefont {M.~R.}\ \bibnamefont {Oduncu}}, \bibinfo {author} {\bibfnamefont {Q.}~\bibnamefont {Yang}}, \bibinfo {author} {\bibfnamefont {S.}~\bibnamefont {Sedaghat}}, \bibinfo {author} {\bibfnamefont {K.}~\bibnamefont {Maize}}, \bibinfo {author} {\bibfnamefont {J.~P.}\ \bibnamefont {Allebach}}, \bibinfo {author} {\bibfnamefont {A.}~\bibnamefont {Shakouri}}, \bibinfo {author} {\bibfnamefont {N.}~\bibnamefont {Glassmaker}},  \emph {et~al.},\ }\bibfield  {title} {\enquote {\bibinfo {title} {Steady-state and transient performance of ion-sensitive electrodes suitable for wearable and implantable electro-chemical sensing},}\ }\href@noop {} {\bibfield  {journal} {\bibinfo  {journal} {IEEE Transactions on Biomedical Engineering}\ }\textbf {\bibinfo {volume} {69}},\ \bibinfo {pages} {96--107}
  (\bibinfo {year} {2021})}\BibitemShut {NoStop}%
\bibitem [{\citenamefont {Bobacka}, \citenamefont {Lewenstam},\ and\ \citenamefont {Ivaska}(2001)}]{bobacka2001}%
  \BibitemOpen
  \bibfield  {author} {\bibinfo {author} {\bibfnamefont {J.}~\bibnamefont {Bobacka}}, \bibinfo {author} {\bibfnamefont {A.}~\bibnamefont {Lewenstam}}, \ and\ \bibinfo {author} {\bibfnamefont {A.}~\bibnamefont {Ivaska}},\ }\bibfield  {title} {\enquote {\bibinfo {title} {Equilibrium potential of potentiometric ion sensors under steady-state current by using current-reversal chronopotentiometry},}\ }\href {\doibase https://doi.org/10.1016/S0022-0728(00)00515-5} {\bibfield  {journal} {\bibinfo  {journal} {Journal of Electroanalytical Chemistry}\ }\textbf {\bibinfo {volume} {509}},\ \bibinfo {pages} {27--30} (\bibinfo {year} {2001})}\BibitemShut {NoStop}%
\bibitem [{\citenamefont {Jin}\ \emph {et~al.}(2017)\citenamefont {Jin}, \citenamefont {Jiang}, \citenamefont {Song}, \citenamefont {Fang}, \citenamefont {Rogers},\ and\ \citenamefont {Alam}}]{jin2017}%
  \BibitemOpen
  \bibfield  {author} {\bibinfo {author} {\bibfnamefont {X.}~\bibnamefont {Jin}}, \bibinfo {author} {\bibfnamefont {C.}~\bibnamefont {Jiang}}, \bibinfo {author} {\bibfnamefont {E.}~\bibnamefont {Song}}, \bibinfo {author} {\bibfnamefont {H.}~\bibnamefont {Fang}}, \bibinfo {author} {\bibfnamefont {J.~A.}\ \bibnamefont {Rogers}}, \ and\ \bibinfo {author} {\bibfnamefont {M.~A.}\ \bibnamefont {Alam}},\ }\bibfield  {title} {\enquote {\bibinfo {title} {Stability of mosfet-based electronic components in wearable and implantable systems},}\ }\href {\doibase 10.1109/TED.2017.2715837} {\bibfield  {journal} {\bibinfo  {journal} {IEEE Transactions on Electron Devices}\ }\textbf {\bibinfo {volume} {64}},\ \bibinfo {pages} {3443--3451} (\bibinfo {year} {2017})},\ \bibinfo {note} {doi: \href{https://doi.org/10.1109/TED.2017.2715837}{10.1109/TED.2017.2715837}}\BibitemShut {NoStop}%
\bibitem [{\citenamefont {Sze}\ and\ \citenamefont {Ng}(2006)}]{sze2006}%
  \BibitemOpen
  \bibfield  {author} {\bibinfo {author} {\bibfnamefont {S.~M.}\ \bibnamefont {Sze}}\ and\ \bibinfo {author} {\bibfnamefont {K.~K.}\ \bibnamefont {Ng}},\ }\href@noop {} {\emph {\bibinfo {title} {Physics of semiconductor devices}}}\ (\bibinfo  {publisher} {John Wiley \& Sons},\ \bibinfo {year} {2006})\BibitemShut {NoStop}%
\bibitem [{\citenamefont {Bergveld}(1991)}]{Bergveld1991}%
  \BibitemOpen
  \bibfield  {author} {\bibinfo {author} {\bibfnamefont {P.}~\bibnamefont {Bergveld}},\ }\bibfield  {title} {\enquote {\bibinfo {title} {{A critical evaluation of direct electrical protein detection methods}},}\ }\href {\doibase 10.1016/0956-5663(91)85009-L} {\bibfield  {journal} {\bibinfo  {journal} {Biosensors and Bioelectronics}\ }\textbf {\bibinfo {volume} {6}},\ \bibinfo {pages} {55--72} (\bibinfo {year} {1991})}\BibitemShut {NoStop}%
\bibitem [{\citenamefont {Amemiya}, \citenamefont {B{\"u}hlmann},\ and\ \citenamefont {Odashima}(2003)}]{Amemiya2003}%
  \BibitemOpen
  \bibfield  {author} {\bibinfo {author} {\bibfnamefont {S.}~\bibnamefont {Amemiya}}, \bibinfo {author} {\bibfnamefont {P.}~\bibnamefont {B{\"u}hlmann}}, \ and\ \bibinfo {author} {\bibfnamefont {K.}~\bibnamefont {Odashima}},\ }\bibfield  {title} {\enquote {\bibinfo {title} {{A Generalized Model for Apparently ``Non-Nernstian'' Equilibrium Responses of Ionophore-Based Ion-Selective Electrodes. 1. Independent Complexation of the Ionophore with Primary and Secondary Ions}},}\ }\href {\doibase 10.1021/ac026471g} {\bibfield  {journal} {\bibinfo  {journal} {Analytical Chemistry}\ }\textbf {\bibinfo {volume} {75}},\ \bibinfo {pages} {3329--3339} (\bibinfo {year} {2003})},\ \bibinfo {note} {doi: \href{https://doi.org/10.1021/ac026471g}{10.1021/ac026471g}}\BibitemShut {NoStop}%
\bibitem [{\citenamefont {Mele}\ \emph {et~al.}(2022{\natexlab{a}})\citenamefont {Mele}, \citenamefont {Palestri}, \citenamefont {Alam},\ and\ \citenamefont {Selmi}}]{Mele2022b}%
  \BibitemOpen
  \bibfield  {author} {\bibinfo {author} {\bibfnamefont {L.~J.}\ \bibnamefont {Mele}}, \bibinfo {author} {\bibfnamefont {P.}~\bibnamefont {Palestri}}, \bibinfo {author} {\bibfnamefont {M.~A.}\ \bibnamefont {Alam}}, \ and\ \bibinfo {author} {\bibfnamefont {L.}~\bibnamefont {Selmi}},\ }\bibfield  {title} {\enquote {\bibinfo {title} {Selectivity, sensitivity and detection range in ion-selective membrane-based electrochemical potentiometric sensors analyzed with poisson-boltzmann equilibrium model},}\ }\href {\doibase 10.1109/JSEN.2022.3185168} {\bibfield  {journal} {\bibinfo  {journal} {IEEE Sensors Journal}\ }\textbf {\bibinfo {volume} {22}},\ \bibinfo {pages} {15010--15021} (\bibinfo {year} {2022}{\natexlab{a}})}\BibitemShut {NoStop}%
\bibitem [{\citenamefont {Jasielec}\ \emph {et~al.}(2012)\citenamefont {Jasielec}, \citenamefont {Filipek}, \citenamefont {Szyszkiewicz}, \citenamefont {Fausek}, \citenamefont {Danielewski},\ and\ \citenamefont {Lewenstam}}]{jasielec2012}%
  \BibitemOpen
  \bibfield  {author} {\bibinfo {author} {\bibfnamefont {J.}~\bibnamefont {Jasielec}}, \bibinfo {author} {\bibfnamefont {R.}~\bibnamefont {Filipek}}, \bibinfo {author} {\bibfnamefont {K.}~\bibnamefont {Szyszkiewicz}}, \bibinfo {author} {\bibfnamefont {J.}~\bibnamefont {Fausek}}, \bibinfo {author} {\bibfnamefont {M.}~\bibnamefont {Danielewski}}, \ and\ \bibinfo {author} {\bibfnamefont {A.}~\bibnamefont {Lewenstam}},\ }\bibfield  {title} {\enquote {\bibinfo {title} {{Computer simulations of electrodiffusion problems based on Nernst--Planck and Poisson equations}},}\ }\href {\doibase 10.1016/j.commatsci.2012.05.054} {\bibfield  {journal} {\bibinfo  {journal} {Computational materials science}\ }\textbf {\bibinfo {volume} {63}},\ \bibinfo {pages} {75--90} (\bibinfo {year} {2012})}\BibitemShut {NoStop}%
\bibitem [{\citenamefont {Mele}\ \emph {et~al.}(2022{\natexlab{b}})\citenamefont {Mele}, \citenamefont {Palestri}, \citenamefont {Selmi},\ and\ \citenamefont {Alam}}]{Mele2022a}%
  \BibitemOpen
  \bibfield  {author} {\bibinfo {author} {\bibfnamefont {L.~J.}\ \bibnamefont {Mele}}, \bibinfo {author} {\bibfnamefont {P.}~\bibnamefont {Palestri}}, \bibinfo {author} {\bibfnamefont {L.}~\bibnamefont {Selmi}}, \ and\ \bibinfo {author} {\bibfnamefont {M.~A.}\ \bibnamefont {Alam}},\ }\bibfield  {title} {\enquote {\bibinfo {title} {Modeling non-equilibrium ion-transport in ion-selective-membrane/electrolyte interfaces for electrochemical potentiometric sensors},}\ }\href {\doibase 10.1109/JSEN.2022.3178297} {\bibfield  {journal} {\bibinfo  {journal} {IEEE Sensors Journal}\ }\textbf {\bibinfo {volume} {22}},\ \bibinfo {pages} {12987--12996} (\bibinfo {year} {2022}{\natexlab{b}})}\BibitemShut {NoStop}%
\bibitem [{\citenamefont {Alam}\ and\ \citenamefont {Lundstrom}(1994)}]{Alam1994}%
  \BibitemOpen
  \bibfield  {author} {\bibinfo {author} {\bibfnamefont {M.}~\bibnamefont {Alam}}\ and\ \bibinfo {author} {\bibfnamefont {M.}~\bibnamefont {Lundstrom}},\ }\bibfield  {title} {\enquote {\bibinfo {title} {Simple analysis of carrier transport and buildup in separate confinement heterostructure quantum well lasers},}\ }\href {\doibase 10.1109/68.392225} {\bibfield  {journal} {\bibinfo  {journal} {IEEE Photonics Technology Letters}\ }\textbf {\bibinfo {volume} {6}},\ \bibinfo {pages} {1418--1420} (\bibinfo {year} {1994})}\BibitemShut {NoStop}%
\bibitem [{\citenamefont {Jin}\ \emph {et~al.}(2020)\citenamefont {Jin}, \citenamefont {Bandodkar}, \citenamefont {Fratus}, \citenamefont {Asadpour}, \citenamefont {Rogers},\ and\ \citenamefont {Alam}}]{Jin2020}%
  \BibitemOpen
  \bibfield  {author} {\bibinfo {author} {\bibfnamefont {X.}~\bibnamefont {Jin}}, \bibinfo {author} {\bibfnamefont {A.~J.}\ \bibnamefont {Bandodkar}}, \bibinfo {author} {\bibfnamefont {M.}~\bibnamefont {Fratus}}, \bibinfo {author} {\bibfnamefont {R.}~\bibnamefont {Asadpour}}, \bibinfo {author} {\bibfnamefont {J.~A.}\ \bibnamefont {Rogers}}, \ and\ \bibinfo {author} {\bibfnamefont {M.~A.}\ \bibnamefont {Alam}},\ }\bibfield  {title} {\enquote {\bibinfo {title} {Modeling, design guidelines, and detection limits of self-powered enzymatic biofuel cell-based sensors},}\ }\href {\doibase https://doi.org/10.1016/j.bios.2020.112493} {\bibfield  {journal} {\bibinfo  {journal} {Biosensors and Bioelectronics}\ }\textbf {\bibinfo {volume} {168}},\ \bibinfo {pages} {112493} (\bibinfo {year} {2020})}\BibitemShut {NoStop}%
\end{thebibliography}%

\end{document}